\documentclass{article}
\usepackage{amsmath}
\usepackage{amsfonts}
\usepackage{amssymb}
\usepackage{graphicx}
\newcommand{\be}{\begin{equation}}
\newcommand{\ee}{\end{equation}}

\newcommand{\bea}{\begin{eqnarray}}
\newcommand{\eea}{\end{eqnarray}}

\newcommand{\al}{\alpha}
\newcommand{\bt}{\beta}
\newcommand{\gm}{\gamma}

\newcommand{\dl}{\delta}

\newcommand{\eps}{\epsilon}

\newcommand{\et}{\eta}

\newcommand{\kp}{\kappa}
\newcommand{\lm}{\lambda}

\newcommand{\rh}{\rho}

\newcommand{\sg}{\sigma}

\newcommand{\ta}{\tau}

\newcommand{\ph}{\phi}

\newcommand{\om}{\omega}

\newcommand{\rarrow}{\rightarrow}
\newcommand{\Rarrow}{\Rightarrow}

\newcommand{\nn}{\nonumber}

\newcommand{\varep}{\varepsilon}

\begin{document}

\begin{center}

\Large{\bf Generating Solutions to the Einstein - Maxwell Equations}\\

\vspace{1.cm}

\large{I. G. Contopoulos$^1$, F. P. Esposito$^2$, K. Kleidis$^3$,}\\
\large{D. B. Papadopoulos$^4$, and L. Witten$^5$}\\

\vspace{.5cm}

{\small $^1$Research Center for Astronomy and Applied Mathematics,}\\
{\small Academy of Athens, 115 27 Athens, Greece}\\
{\small $^2$Department of Physics, University of Cincinnati, 45269 Ohio, USA}\\
{\small $^3$Department of Mechanical Engineering, Technological Education Institute of Central Macedonia, 621 24 Serres, Greece}\\
{\small $^4$Department of Physics, Aristotle University of Thessaloniki,}\\
{\small 541 24 Thessaloniki, Greece} \\
{\small $^5$Department of Physics, University of Florida, 32611-8440 Florida, USA}

\end{center}


\bigskip

\begin{abstract}

The Einstein-Maxwell (E-M) equations in a curved spacetime that admits at least one Killing vector are derived, from a Lagrangian density adapted to symmetries. In this context, an auxiliary space of potentials is introduced, in which, the set of potentials associated to an original (seed) solution of the E-M equations are transformed to a new set, either by continuous transformations or by discrete transformations. In this article, continuous transformations are considered. Accordingly, originating from the so-called $\gm_A$-metric, other exact solutions to the E-M equations are recovered and discussed.

\end{abstract}

\section{Introduction}

Until the early 70s, exact solutions to the Einstein field equations were notoriously difficult to be obtained. Since then, however, many interesting solutions have been found, upon the exploitation of curved spacetimes with symmetries, i.e., manifolds admitting Killing vectors. Geroch~\cite{1} was one of the first researchers, who systematically employed symmetries to produce new classes of solutions.

On the other hand, many solutions to the E-M equations are already known~\cite{2}, and some of them, like the Reissner-Nordstr{\"o}m solution or/and the Kerr-Newman solution, have played an important role in the development of many areas of Astrophysics. Detailed reviews of known solutions, and their classification schemes, can be found in the work of Ehlers and Kundt~\cite{3}, Kramer, Neugebauer and Stephani~\cite{4}, Kinnersley~\cite{5}, \cite{6} and Petrov~\cite{7}. More recently, new asymptotically-flat solutions to the Ernst equations for the E-M system were found by Manko et. al.~\cite{8}, while particular solutions to the Einstein field equations and their role in General Relativity (GR) and Astrophysics have been discussed by Bicak~\cite{9}.

Today, there are several techniques suggesting how to generate solutions to the Einstein and the E-M equations, originating from already existing ones~\cite{2}. Richterek et al.~\cite{10},~\cite{11}, in particular, have developed a technique that is based on the striking analogy between the equations satisfied by the Killing vectors of a vacuum solution and the corresponding (sourceless) Maxwell equations. The so called Horsky-Mitskievich conjecture, outlines an efficient and fruitful way to obtain solutions to the E-M equations, as a generalization of some already known vacuum seed metrics. In this context, taking the $\gm$-solution as a seed vacuum spacetime, they have obtained two classes of E-M fields, the main properties of which are discussed extensively in~\cite{10} and~\cite{11}.

The analytic description of the curved spacetime surrounding a realistic astrophysical object still remains an open problem, although many approaches have been attempted by means of numerical techniques (see, e.g.,~\cite{12}). In this context, the existence of a consistent, analytic representation of the vacuum metric outside the astrophysical object under consideration is desirable for several reasons, e.g., the computation becomes simpler, the study of dynamical properties of the curved spacetime (such as gravitational radiation) is possible, etc. Although there has been a remarkable progress in finding exact solutions to the E-M equations, still there are not enough solutions with sources that satisfy realistic physical requirements.

In this article we discuss spacetime models with symmetries, i.e., spacetimes that admit at least one Killing vector field. The main scope of the article is to describe, in a unified way, methods of deriving families of solutions to the E-M equations, originating from already known solutions that admit symmetries. To do so, we use the effective formalism developed by Geroch~\cite{1}, and the method used by Neugebauer and Kramer~\cite{4}, \cite{13}. It is worth noting that, the solution generating transformations employed in the technique we intend to analyse, were independently discussed also by Harisson~\cite{14} and Kinnersley~\cite{5},~\cite{6}.

The article is organized as follows: In Section 2, we outline the Geroch formalism that led to the method of generating solutions by exploiting symmetries in the (auxiliary) potential space, and in Section 3, we review the method of Neugebauer and Kramer that led to Kinnersley transformations. As an application of this method, we use the Kinnersley V transformation upon a spacelike Killing vector of the $\gm_A$-metric, to generate an exact solution to the E-M equations, that belongs to the broad class of magnetized solutions given by Eq. (10) of Richterek et al.~\cite{11}. Finally, in Section 4, we use the canonical form of the $\gm_A$-solution, to generate more exact solutions to the E-M equations, using either a timelike or a spacelike Killing vector. As we demonstrate, each and everyone of these solutions describe stationary electromagnetic (e/m) fields in vacuum. We conclude in Section 5.

\section{The E-M equations in spacetimes admitting symmetries}

In this Section, we outline the formalism developed (mainly) by Geroch~\cite{1}. Accordingly, we derive the E-M equations on a spacetime manifold $(M, \: g_{\al \bt})$ that admits symmetries, related to a Killing vector field, $\xi^{\al}$ (Greek indices refer to the four-dimensional spacetime). In what follows, we take this Killing vector to be timelike, $\xi_{\al} \xi^{\al} = - \lm$, where $(- \lm)$ is the negative norm of $\xi^{\al}$. If $\xi^{\al}$ was taken to be spacelike, several signs in what follows would have to be changed.

Let $S$ be the three-dimensional manifold of the trajectories of the Killing vector field $\xi^{\al}$, i.e., any point on $S$ is a trajectory of $\xi^{\al}$. Such a manifold necessarily inherits a differential structure, being the quotient manifold of $M$ under the action of the Lie group that generates $\xi^{\al}$. Any function, $f$, defined on $S$, corresponds to a function of $M$ that remains constant along the trajectories of $\xi^{\al}$, i.e., ${\cal L}_{\xi} f = 0$, where ${\cal L}_{\xi}$ is the Lie derivative along the vector field $\xi^{\al}$. Correspondingly, any tensor defined on $S$, $T^{a ... b}_{c ... d}$ (Latin indices refer to the three-dimensional quotient manifold), corresponds to a tensor of $M$, $T^{\al ... \bt}_{\gm ... \dl}$, that \texttt{(a)} remains constant along the trajectories of $\xi^{\al}$, i.e., ${\cal L}_{\xi} T^{\al ... \bt}_{\gm ... \dl} = 0$ and, in addition, \texttt{(b)} has zero projection in that direction, i.e., $\xi^{\gm} T^{\al ... \bt}_{\gm ... \dl} = 0 , \: .... \: , \xi_{\bt} T^{\al ... \bt}_{\gm ... \dl} = 0$ (contraction on any index). In fact, correspondence between tensors on $S$ and the associated quantities of $M$ that satisfy conditions \texttt{(a)} and \texttt{(b)} is one-to-one. Accordingly, the procedure is to reduce all the equations valid on the four-dimensional spacetime $M$, to relationships on the three-dimensional manifold $S$. Recall that, since $\xi^{\al}$ is a Killing vector, we have \be {\cal L}_{\xi} g_{\al \bt} = 0 = 2 \nabla _{(\al} \xi_{\bt)} \ee and \be \nabla _{[\al} \nabla _{\bt]} \xi_{\gm} = \xi^{\dl} {\cal R}_{\dl \al \bt \gm} \: . \ee In Eqs. (1) and (2), $g_{\al \bt}$ is the metric tensor of signature $+2$ attributed to $M$, $\nabla_{\al}$ denotes covariant derivative on $M$, ${\cal R}_{\dl \al \bt \gm}$ is the corresponding Riemann tensor, with contractions ${\cal R}_{\al \gm} = g^{\dl \bt} {\cal R}_{\dl \al \bt \gm}$ (Ricci tensor) and ${\cal R} = g^{\al \gm} {\cal R}_{\al \gm}$ (scalar curvature), while the symbols $_{(\al} \nabla _{\bt)}$ and $_{[\al} \nabla _{\bt]}$ stand for symmetric and antisymmetric differentiation, respectively.

Upon consideration of a timelike Killing vector, $\xi^{\al}$, on $M$, we take $S$ to be the three-dimensional slice that is perpendicular to $\xi^{\al}$. Accordingly, we define a symmetric non-degenerate tensor field on $S$, of signature $+3$, $h_{ab} = g_{ab} + \lm ^{-1} \xi_a \xi_b$, and, therefore, a Riemannian metric on this space. Clearly, if $\xi^{\al}$ was taken to be spacelike, then $h_{ab}$ would have been pseudo-Riemannian.

Now, let $T^{b ... c}_{d ... e}$ be a tensor field on $S$. The corresponding covariant derivative is defined by \be D_a T^{b ... c}_{d ... f} = h^m_a \left ( h^n_d ... h^o_f \right ) \left ( h^b_p ... h^c_q \right ) \nabla_m T^{p ... q}_{n ... o} \: . \ee Clearly, $D_a T^{b ... c}_{d ... e}$ is a tensor field on $S$. As a consequence, $(S, \: h_{ab})$ is a Riemannian manifold with Riemannian connection $D_a$, something that stems from the fact that, $D_a h_{bc} = 0$, and, in addition, $D _{[a} D_{b]} f = 0$ for all functions on $S$, since $\nabla_a$ is torsion free.

Moreover, on M, we define the twist vector, $\om^{\al}$, associated to the Killing vector $\xi^{\al}$, as \be \om^{\al} = \frac{1}{2} \et^{\al \bt \gm \dl} \xi_{\bt} \nabla_{\gm} \xi_{\dl} \: , \ee in terms of which we obtain \be \nabla_{\al} \left ( \xi_{\bt} \right ) = - \lm^{-1} \xi_{[\al} \nabla_{\bt]} \lm - \lm^{-1} \et_{\al \bt \gm \dl} \xi^{\gm} \om^{\dl} \: . \ee In Eqs. (4) and (5), $\et_{\al \bt \gm \dl} = \left ( \det \vert \vert g_{\mu \nu} \vert \vert \right )^{1/2} \varep_{\al \bt \gm \dl}$ is the alternating tensor on four-dimensions, where $\varep_{\al \bt \gm \dl}$ is the completely antisymmetric symbol of four indices, with $\varep_{0 1 2 3} = 1$. Since $\xi_{\al} \om^{\al} = 0 = {\cal L}_{\xi} \om^a$, it follows that $\om^{\al}$ is a vector also on $S$.

Now, given $g_{\al \bt}$, we can determine $\xi^{\al}$, and through that, $\lm$, $h_{ab}$ and $\om^{\al}$, as well. Reversely, upon consideration of $\lm$, $\xi^{\al}$, $h_{ab}$, and $\om^{\al}$, it is possible to reconstruct the metric $g_{\al \bt}$ of $M$ (see, e.g.,~\cite{15}). Notice that, since $\om^{\al}$ is a vector also on $S$, indices can be raised either by $h^{ab}$ or by $g^{\al \bt}$. Accordingly, Eq. (5) is equivalent to \be \nabla_{[ \al} \lm \xi_{\bt ]} = - \et_{\al \bt \gm \dl} \xi^{\gm} \om^{\dl}\: . \ee Integration of Eq. (6) yields the quantity $\lm \xi_{\bt}$ up to a gradient (equivalent to a coordinate transformation) and therefore, in principle, both $\xi_{\bt}$ and the metric $g_{\al \bt}$ can be determined.

From Eq. (4), we furthermore obtain \be \nabla _{[ \al} \om_{\bt ]} = - \frac{1}{2} \et_{\al \bt \gm \dl} \xi^{\gm} {\cal R}_{\mu}^{\dl} \xi^{\mu} \: , \ee in view of which, if ${\cal R}_{\al \bt} = 0$ or ${\cal R}_{\al \bt} \xi^{\bt} \propto \xi_{\al}$, then, locally, there exists a scalar, $\om$, such that $\om_{\al} = \nabla_{\al} \om$. As we will show later on, for stationary e/m fields, Eq. (7) results in an important simplification of the E-M equations.

On $S$, the equations that determine the scalar $\lm$ and the vector $\om^a$ may be derived in an analogous fashion. In fact, by analogy to Eq. (7), we have \be D_{[a} \om_{b]} = - \frac{1}{2} \et_{abcd} \xi^c {\cal R}_m^d \xi^m \: . \ee Now, in order to determine the divergence of $\om^a$ on $S$, $D_a \om^a$, we use the general definition given by Eq. (3), i.e., \be D_a \om^a = h_c^b \nabla_b \om^c \: , \ee where the divergence of $\om^{\al}$ on $M$ is given by \be \nabla_{\al} \omega^{\al} = \frac{1}{2} \et_{\al \bt \gm \dl} \left ( \nabla^{\al} \xi^{\bt} \right ) \left ( \nabla^{\gm} \xi^{\dl} \right ) \: , \ee since \be {\cal R}^{\al [ \bt \gm \dl ]} = 0 \: . \ee Upon consideration of Eq. (5) with $\al = 1, 2, 3 = a$, Eq. (10) yields \be \nabla_a \om^a = 2 \lm^{-1} \om_b D^b \lm \ee and, therefore, Eq. (9) results in  \be D_a \om^a = h_c^b \nabla_b \om^c = \nabla_b \om^b - \frac{1}{2}\lm^{-1} \om^e D_e \lm = \frac{3}{2} \lm^{-1} \om^b D_b \lm \ee or, equivalently, \be D_a \left ( \lm^{- \frac{3}{2}} \om^a \right ) = 0 \: . \ee

Moreover, on the four-dimensional manifold $M$, the wave operator (d' Alembertian) of $\lm$ is given by \be \nabla^{\al} \nabla_{\al} \lm = \lm^{-1} \left ( \nabla_{\al} \lm \right ) \left ( \nabla^{\al} \lm \right ) - 4 \lm^{-1} \om^{\al} \om_{\al} + 2 {\cal R}^{(M)}_{\al \bt} \xi^{\al} \xi^{\bt} \: , \ee the proof of which is given in the Appendix A. Similarly, on the quotient manifold $S$, the d' Alembertian of $\lm$ reads \be D^a D_a \lm = \frac{1}{2} \lm^{-1} \left ( D_a \lm \right ) \left ( D^a \lm \right ) - 4 \lm^{-1} \om^a \om_a + 2 {\cal R}^{(S)}_{ab} \xi^a \xi^b \ee or, else, \be \lm^{\frac{1}{2}} D^a \left ( \lm ^{-\frac{1}{2}} D_a \lm \right ) = - 4 \lm^{-1} \om^a \om_a + 2 {\cal R}^{(S)}_{ab} \xi^a \xi^b \: . \ee Now, the derivation of the Ricci tensor on $S$, ${\cal R}^{(S)}_{bd}$, in terms of $\lm$ and $\om^a$, is straightforward, and is given in the Appendix B. Accordingly, ${\cal R}^{(S)}_{bd}$ in terms of the Ricci tensor on $M$, ${\cal R}^{(M)}_{\al \bt}$, is given by \bea {\cal R}^{(S)}_{bd} & = & h_b^q h_d^s {\cal R}^{(M)}_{qs} + \frac {\lm^{-1}}{2} D_b D_d \lm - \frac {\lm^{-2}}{4} \left ( D_b \lm \right ) \left ( D_d \lm \right ) \nn \\ & + & 2 \lm^{-2} \left ( \om_b \om_d - h_{bd} \om_a \om^a \right ) \: . \eea Since ${\cal R}^{(M)}_{\al \bt}$ satisfies the Einstein field equations on $M$, by virtue of Eq. (18), Eqs. (13) and (16) are equivalent to the Einstein equations on $S$. These equations can be simplified significantly upon a conformal transformation of $h_{ab}$, of the form \be \tilde{h}_{ab} = \lm h_{ab} \: , \ee in terms of which, the Ricci tensor on $S$ transforms to \be \tilde{\cal R}^{(S)}_{ab} = \frac{1}{2} \lm^{-2} \left [ 4 \om_a \om_b + \left ( D_a \lm \right ) \left ( D_b \lm \right ) \right ] + h_a^c h_b^d  \left [ {\cal R}^{(S)}_{cd} - \lm^{-1} h_{cd} {\cal R}^{(S)}_{mn} \xi^m \xi^n \right ] \: . \ee Now, in view of the Einstein equations on $M$ $(8 \pi G = 1 = c)$, \be {\cal R}^{(M)}_{\al \bt} - \frac{1}{2} g_{\al \bt} {\cal R}^{(M)} = {\cal T}^{(M)}_{\al \bt} \: , \ee where ${\cal T}^{(M)}_{\al \bt}$ is the corresponding energy-momentum tensor, Eqs. (13) and (16) are written in the form \be \tilde{D}^a G_a = \tilde{h}^{ab} \left ( G_a - G_a^* \right ) G_b + \lm ^{-2} \left ( {\cal T}^{(M)}_{ab} \xi^a \xi^b + \frac{\lm}{2} {\cal T} \right ) \ee and \be \tilde{\cal R}^{(S)}_{ab} = 2 G_{(a} {G^*}_{b)} + \tilde{h}_a^c \tilde{h}_b^d \left [ {\cal T}^{(M)}_{cd} - \lm^{-2} h_{cd} \left ( {\cal T}^{(M)}_{mn} \xi^m \xi^n \right ) \right ] \: , \ee where ${\cal T}$ is the trace of the energy-momentum tensor, ${\cal T}^{(M)}_{\al \bt}$, and we have set \be G_a = \frac{1}{2} \lm^{-1} \left ( \tilde{D}_a \lm + 2 \imath \om_a \right ) \: , \ee with $\tilde{D}_a $ being the covariant derivative on $S$ with respect to the conformal metric $\tilde{h}_{ab}$. In vacuum spacetime, i.e., as long as ${\cal T}^{(M)}_{\al \bt} = 0$, Eqs. (22) and (23), reduce to \be \tilde{D}^a G_a = \tilde{h}^{ab} \left ( G_a - G_a^* \right ) G_b \ee and \be \tilde{\cal R}^{(S)}_{ab} = 2 G_{(a} {G^*}_{b)} \: , \ee respectively. Instead of analyzing these equations, we will now apply the aforementioned formalism to stationary e/m fields, in order to demonstrate that the E-M equations can take a form similar to Eqs. (25) and (26).

Let ${\cal F}_{\al \bt}$ be the Faraday tensor of the e/m field in the four-dimensional spacetime $\left ( M, \: g_{\al\bt} \right )$, which admits (at least) one Killing vector, $\xi^{\bt}$. We define by \be E_{\al} = {\cal F}_{\al \bt} \xi^{\bt} ~~ \mbox{and}~~~ B_{\al} = \frac{1}{2} \et_{\al \bt \gm \dl} \xi^{\bt} {\cal F}^{\gm \dl} \: , \ee the electric and the magnetic component, respectively, of ${\cal F}_{\al \bt}$ along the direction of $\xi^{\bt}$. By virtue of $E_{\al}$ and $B_{\al}$, ${\cal F}_{\al \bt}$ can be written in the form \be {\cal F}_{\al \bt} = 2 \lm^{-1} \xi _{[ \al} E_{\bt ]} - \lm^{-1} \et_{\al \bt \gm \dl} \xi^{\gm} B^{\dl} \: . \ee As far as stationary e/m fields are concerned, the Maxwell equations in vacuum are written in the form \be {\cal L}_{\xi} {\cal F}_{\al \bt} = 0 \: \Rarrow \: D_{[a} E _{b]} = 0 = D_{[a} B_{b]} \: . \ee In this case, locally, there exist two potentials, $\eps$ and $\bt$, attributed to $E_a$ and $B_a$, respectively, defined by \be E_a = D_a \eps ~~~\mbox{and}~~~~ B_a = D_a \bt \: . \ee In terms of $\eps$ and $\bt$, Eqs. (29) are written in the form \be D^a \left ( \lm^{-\frac{1}{2}} D_a \eps \right ) = 2 \lm^{- \frac{3}{2}} \om^a D_a \bt \: ,~~~ \mbox{and}~~~~ D^a \left ( \lm^{-\frac{1}{2}} D_a \bt \right ) = - 2 \lm^{-\frac{3}{2}} \om^a D_{a} \eps \: . \ee Accordingly, we define \be \Psi = \eps + \imath \bt \: , \ee in terms of which Eqs. (31) reduce to \be D^a \left ( \lm^{-\frac{1}{2}} D_a \Psi \right ) = 2 \imath \lm^{-\frac{3}{2}} \om^a D_a \Psi \: . \ee

In general, $\om^a$ is not curl free. Nevertheless, in the presence of stationary e/m fields in vacuum, a curl free vector does exist, namely, \be \psi_a \equiv D_a \psi = \om_a + \frac {\imath}{4} \left ( \Psi D_a \Psi^* - \Psi^* D_a \Psi \right ) \: . \ee The fact that $\psi_a$ is curl free can be directly deduced from Eqs. (13), (16) and (32). In view of all the above, we now introduce the Ernst potential~\cite{16}, \be {\cal E} = \lm - \frac{1}{2} \Psi \Psi^* + 2 \imath \psi \: , \ee for which, by virtue of Eq. (34), we have \be \tilde{D}_a {\cal E} = 2 \lm G_a - \Psi \tilde{D}_a \Psi^* \: . \ee In view of Eq. (36), the d' Alembertian of ${\cal E}$ in terms of the conformal metric $\tilde{h}_{ab}$ reads \be \tilde{D}^a \tilde{D}_a {\cal E} = 2 \lm \tilde{D}^a G_a + 2 \left ( \tilde{D}^a \lm \right ) G_a - \left ( \tilde{D}^a \Psi \right ) \left ( \tilde{D}_a \Psi^* \right ) - \Psi \tilde{D}^a \tilde{D}_a \Psi^* \: , \ee which, upon consideration of Eqs. (25), (33) and (35), results in \be \lm \tilde{D}^a \tilde{D}_a {\cal E} = \tilde{h}^{ab} \tilde{D}_a {\cal E} \left ( \tilde{D}_b {\cal E} + \Psi \tilde{D}_b \Psi ^* \right ) \: . \ee

Hence, the complete set of the E-M equations for stationary, electrovacuum spacetimes is summarized as follows:\\

\textbf{GEOMETRY:} \bea {\cal L}_{\xi} \: g_{\al \bt} \: = & 0 & = \: {\cal L}_{\xi} \: {\cal F}_{\al \bt} \: , \nn \\ \xi^{\al} \xi_{\al} & = & - \: \lm \: , \nn \\ \om^{\al} = & \frac{1}{2} & \et^{\al \bt \gm \dl} \xi_{\bt} \nabla_{\gm} \xi_{\dl} \: , \nn \\ \tilde{h}_{ab} & = & \lm g_{ab} + \xi_a \xi_b \: . \eea

\textbf{POTENTIALS:} \bea \Psi & = & \eps + \imath \bt \: , \nn \\ \tilde{D}_a \psi & = & \om_a + \frac{\imath}{4} \left ( \Psi \tilde{D}_a \Psi^* - \Psi^* \tilde{D}_a \Psi \right ) \: , \nn \\ {\cal E} & = & \lm - \frac{1}{2} \Psi \Psi^* + 2 \imath \psi \: . \eea

\textbf{FIELD EQUATIONS:} \bea \lm \tilde{D}^a \tilde{D}_a {\cal E} & = & \tilde{h}^{ab} \left ( \tilde{D}_a {\cal E} + \Psi \tilde{D}_a \Psi^* \right ) \tilde{D}_b {\cal E} \: , \nn \\ \lm \tilde{D}^a \tilde{D}_a \Psi & = & \tilde{h}^{ab} \left ( \tilde{D}_a {\cal E}^* + \Psi^* \tilde{D}_a \Psi \right ) \tilde{D}_b \Psi \eea and \bea 2 \lm^2 {\cal R}^{(S)}_{ab} & = & \tilde{D} _{(a} {\cal E} \tilde{D} _{b)} {\cal E}^* + \Psi^* \tilde{D} _{(a} {\cal E} \tilde{D} _{b)} \Psi + \Psi \tilde{D} _{(a} \Psi^* \tilde{D} _{b)} {\cal E}^* \nn \\ & - & \left ( {\cal E} + {\cal E}^* \right ) \tilde{D} _{(a} \Psi \tilde{D} _{b)} \Psi^* \: . \eea

The field equations (41) and (42) can be derived also by means of Lagrangian formalism. Accordingly, upon consideration of Eq. (26), the Einstein-Hilbert Langrangian on $M$, in terms of the various $S$-quantities, is written in the form \be \sqrt{-g} \: \: {\cal R}^{(M)} = \sqrt{\tilde{h}} \: \left \lbrace \tilde{\cal R}^{(S)} - \frac{1}{2} \lm^{-1} \tilde{h}^{ab} \left [ \left ( D_a \lm \right ) \left ( D_b \lm \right ) + 4 \om_a \om_b \right ] + 2 \lm^{-2} {\cal R}^{(S)}_{mn} \xi^m \xi^n \right \rbrace \: , \ee where, in particular, the determinant of the metric tensor on $M$, $g$, is decomposed in terms of the corresponding quantity of the conformal metric on $S$, $\tilde{h}$, as $\sqrt{-g} = \lm^{-1} \sqrt{\tilde{h}}$.

In the case of a stationary electrovacuum space, the rhs of Eq. (43) is written in terms of the potentials as follows \be \sqrt{-g} \: \: {\cal R}^{(M)} =  \sqrt{\tilde{h}} \: \left \lbrace \tilde{\cal R}^{(S)} + \frac{\lm^{-2}}{2} \sqrt{\tilde{h}} \: \tilde{h}^{ab} \left [ 2 \lm \left ( \tilde{D}_a \Psi \right ) \left ( \tilde{D}_a \Psi ^* \right ) - \left ( \tilde{D}_a {\cal E} + \Psi \tilde{D}_a \Psi ^* \right ) \left ( \tilde{D}_b {\cal E}^* + \Psi^* \tilde{D}_b \Psi \right ) \right ] \right \rbrace \: , \ee suggesting that the corresponding Einstein-Hilbert Lagrangian is given by \be L =  \sqrt{\tilde{h}} \left ( \tilde{\cal R}^{(S)} + g_{AB} \Phi^A_{, a} \Phi^{B \: , a} \right ) = \sqrt{\tilde{h}} \: \tilde{h}^{ab} \left ( \tilde{\cal R}^{(S)}_{ab} + g_{AB} \Phi^A_{, a} \Phi^B_{, b} \right ) \: , \ee where the generalized coordinates $\Phi^A$ $(A = 1, 2, 3, 4)$ stand for $\left \lbrace \Phi^1, \: \Phi^2, \: \Phi^3, \: \Phi^4 \right \rbrace \equiv \left \lbrace {\cal E}, \: {\cal E}^*, \: \Psi, \: \Psi^* \right \rbrace$ and $g_{AB}$ is the metric of the four-dimensional space of the potentials. In terms of variations of the Lagrangian (45), Eqs. (41) are equivalent to \be \frac{\dl L}{\dl \Phi^A} = \frac{\dl}{\dl \Phi^A} \left ( \sqrt{\tilde{h}} g_{AB} \Phi^A _{, a} \Phi ^{ B \: , a} \right ) = 0 \: , \ee while Eq. (42) arises from \be \frac{\dl L}{\dl \tilde{h}^{ab}} = 0 \: . \ee

\section{Methods of generating solutions to the E-M equations}

The solution generating method of Neugebauer and Kramer~\cite{13} suggests that, having a solution to the stationary E-M equations (41) and (42), in other words the functional form of $\tilde{h}_{ab}$, $\Psi$, and ${\cal E}$, it is possible to find a new solution, by transforming the original (seed) variables $\left \lbrace \tilde{h}_{ab}, \: \Psi_0, \: {\cal E}_0 \right \rbrace$ into a new set, $\left \lbrace \tilde{h}_{ab}, \: \Psi, \: {\cal E} \right \rbrace$, that leaves the Lagrangian (45) invariant. Under such a transformation, the geometry, $\tilde{h}_{ab}$, of the three-dimensional slice, $S$, also remains invariant, and so does the corresponding Ricci scalar, $\tilde{\cal R}$, as well. Upon consideration of the second term in the Lagrangian (45), such a transformation can be determined as follows.

Let us introduce an auxiliary space, the (generalized) coordinates of which are the (Ernst) potentials; in our case, a real four-dimensional manifold with coordinates $\left \lbrace {\cal E}, \: {\cal E}^*, \: \Psi, \: \Psi^* \right \rbrace$ and metric $g_{AB}$ $\left ( A, \: B = {\cal E}, \: {\cal E}^*, \: \Psi, \: \Psi^* \right )$. Transformations of these coordinates, that leave the Langrangian (45) invariant, are generated by the Killing vectors of this auxiliary (potential) space, $\ell^C$ $\left ( C = {\cal E}, \: {\cal E}^*, \: \Psi, \: \Psi^* \right )$. In other words, first, we solve the Killing equations for the metric $g_{AB}$, \be g_{AB,C} \ell^C + g_{CB} \ell^C_{,A} + g_{AC} \ell^C_{,B} = 0 \: , \ee to determine a Killing vector field of the potential space, $\ell^C = \dot{\Phi}^C = \partial_{\ta} \Phi^C$, where $\ta$ is a length parameter of this space, normalized in the range $0 \leq \ta \leq 1$. Accordingly, we integrate these Killing vectors, to obtain continuous transformations of the generalized coordinates, ${\Phi^{A}}^{\prime} = {\Phi^{A}}^{\prime} \left ( \Phi^B \right )$, that leave the Lagrangian (45) invariant.

The metric tensor of the auxiliary (potential) space is given by \be g_{AB} = \left ( {\cal E} + {\cal E}^* + \Psi \Psi^* \right )^{-2} \left (
\begin{array}{cccc}
0      & 1    & \Psi^*                  & 0                      \\
1      & 0    & 0                       & \Psi                   \\
\Psi^* & 0    & 0                       & -({\cal E}+{\cal E}^*) \\
0      & \Psi & -({\cal E} +{\cal E}^*) & 0
\end{array}
\right ) \ee (see, e.g.,~\cite{17}) and the Killing equations (48) for this metric yield \bea && \ell^{\cal E}_{,{\cal E}^*} + \Psi \ell^{\Psi^*}_{, {\cal E}^*} = 0 \: , \nn \\ && \Psi^* \ell^{\cal E}_{,\Psi} -\left ( {\cal E} + {\cal E}^* \right ) \ell^{\Psi^*}_{,\Psi} = 0 \: , \nn \\ && \ell^{\cal E}_{,\Psi} + \Psi \ell^{\Psi^*}_{,\Psi} + \Psi^* \ell^{\cal E}_{,{\cal E}^*} - \left ( {\cal E} + {\cal E}^* \right ) \ell^{\Psi^*}_{,{\cal E}^*} = 0 \: , \nn \\ && \ell^{\cal E}_{,{\cal E^*}} + \ell^{\cal E}_{,{\cal E}} + \Psi^* \ell^{\Psi}_{,{\cal E^*}} = \frac{ \left ( \ell^{\cal E} + \ell^{{\cal E}^*} + \Psi \ell^{\Psi^*} + \Psi^* \ell^{\Psi} \right )} {2 \left ( {\cal E} + {\cal E}^* + \Psi \Psi^* \right )} \: , \nn \\ && \ell^{\cal E}_{,\Psi} + \Psi^* \left ( \ell^{\cal E}_{{,\cal E}} + \ell^{\Psi}_{,\Psi} \right ) + \ell^{\Psi^*} - \left ( {\cal E} + {\cal E}^* \right ) \ell^{\Psi^*}_{,{\cal E}} \nn \\ & = & \frac{ \left ( \ell^{\cal E} + \ell^{{\cal E}^*} + \Psi \ell^{\Psi^*} + \Psi^* \ell^{\Psi} \right ) \Psi^*} { 2 \left ( {\cal E} + {\cal E}^* + \Psi \Psi^* \right ) } \: , \nn \\ && \Psi^* \ell^{\cal E}_{,\Psi ^*} + \Psi \ell^{{\cal E}^*}_{, \Psi} - \ell^{\cal E} - \ell^{{\cal E}^*} - \left ( {\cal E} + {\cal E}^* \right ) \left ( \ell^{\Psi}_{, \Psi} + \ell^{ \Psi^*}_{, \Psi^*} \right ) \nn \\ & = & \frac{ \left ( \ell^{\cal E} + \ell^{{\cal E}^*} + \Psi \ell^{\Psi^*} + \Psi^* \ell^{\Psi} \right ) \left ( {\cal E} + {\cal E}^* \right )} {2 \left ( {\cal E} + {\cal E}^* + \Psi \Psi^* \right  ) } \: . \eea In each and everyone of the following five classes of solutions to Eqs. (50), first recognized by Kinnersley~\cite{5}, we give the components of the associated Killing vector and then integrate them in the range $0 \leq \ta \leq 1$, to determine the one-parameter group of transformations, that this vector generates. \\

\textbf{Kinnersley I transformation:} \be \ell^{\cal E} = \dot{\cal E} = - \sqrt{2} \al^* \Psi \: ,~~~~\ell^{\Psi} = \dot{\Psi} = \frac{1}{\sqrt{2}} \al \: , \ee where $\al$ is a complex constant. In the range $0 \leq \ta \leq 1$, Eqs. (51) are directly integrated, to give \be \Psi - \Psi_0 = \frac{1}{\sqrt{2}} \: \al \int_0^1 d \ta = \frac{1}{\sqrt{2}} \: \al ~~~\mbox{and}~~~~ {\cal E} = {\cal E}_0 - \sqrt{2} \al^* \Psi_0 - \al \al^* \: . \ee

\textbf{Kinnersley II transformation:} \be \ell^{\cal E} = \dot{\cal E} = \imath \al \: , ~~~~ \ell^{\Psi} = \dot{\Psi} = 0 \: . \ee In this case, the differential equations associated to Eqs. (53) result in \be {\cal E} - {\cal E}_0 = \imath \al \: \int_0^1 d \ta = \imath \al \: , ~~~~ \Psi = \Psi_0 \: . \ee

\textbf{Kinnersley III transformation:} \be \ell^{\cal E} = \dot{\cal E} = - \ln \left ( \bt \bt^* \right ) {\cal E} \: , ~~~ \ell^{\Psi} = \dot{\Psi} = \ln \left ( \frac{\bt}{{\bt^*}^2} \right ) \Psi \: , \ee where $\bt$ is a complex constant. In the same reasoning as above, the associated equations yield the following transformation \be {\cal E} = \left ( \bt \bt^* \right )^{-1} {\cal E}_0 \: , ~~~ \Psi = \frac{\bt}{\left ( \bt^* \right )^2} \Psi_0 \: . \ee

\textbf{Kinnersley IV transformation:} \be \ell^{\cal E} = \dot{\cal E} = - \imath \bt {\cal E}^2 \: , ~~~ \ell^{\Psi} = \dot{\Psi} = \imath \bt {\cal E}^* \Psi \: . \ee In this case, the differential equations associated to Eqs. (57) result in \be {\cal E} = \frac{{\cal E}_0}{1 + \imath \bt {\cal E}_0} \: , ~~~ \Psi = \frac{\Psi_0}{1 + \imath \bt \Psi_0} \: . \ee

\textbf{Kinnersley V transformation:} In this case, the Killing vectors of the potential space are given by \be \ell^{\cal E} = \dot{\cal E} = \sqrt{2} c^* {\cal E} \Psi^* ~~\mbox{and}~~~ \ell^{\Psi} = \dot{\Psi} = \sqrt{2} c^* {\cal E}^* + \sqrt{2} c \Psi^2 \: , \ee where, once again, $c$ is a complex constant. Now, in order to integrate Eqs. (59), we substitute ${\cal E}^* = u^{-1} $ into their complex conjugates, to obtain \be \ddot{u} = - 2 c c^* \: , \ee which, upon integration in the range $0 \leq \ta \leq 1$, results in \be u = \frac{1}{{\cal E}_0^*} + \dl - cc^* \: , \ee where $\dl$ is an integration constant.  Accordingly, \be {\cal E}^* = \frac{{\cal E}_0^*}{1 + \dl {\cal E}_0^* - cc^* {\cal E}_0^* } \: . \ee Now, by virtue of the first of Eqs. (59), we obtain $\dot{{\cal E}^*_0} = \sqrt{2} c {\cal E}^*_0 \Psi_0$, and therefore, \be \dl = - \sqrt{2} c \frac{\Psi_0}{{\cal E}^*_0 } \: . \ee Consequently, the first of the Ernst potentials reads \be {\cal E}^* = \frac{{\cal E}_0^*}{1 - \sqrt{2} c \Psi_0 - cc^* {\cal E}_0^* } \: . \ee On the other hand, in view of Eqs. (59) and (61), we have \be \frac{d {\cal E}^*}{d \Psi} = \frac{{\cal E}^* \Psi}{\frac{c^*}{c} {\cal E}^* + \Psi^2} \: . \ee If we furthermore admit that $ {\cal E}^* = \Phi \Psi^2$, Eq. (65) is written in the form \be \frac{d \Phi}{d \Psi} \Psi^2 + 2 \Phi \Psi = \frac {\Phi \Psi}{1 + \frac{c^*}{c} \Phi} \: , \ee yielding \be \Psi^2 = \left ( \frac{{\cal E}^*}{{\cal E}^*_0} \right )^2 \left ( \Psi^2_0 + 2 \frac{c^*}{c} {\cal E}^*_0 \right ) - 2 \frac{c^*}{c} {\cal E}^* \: . \ee Eventually, the combination of Eqs. (64) and (67) results in \be \Psi = \frac{\Psi_0 + \sqrt{2} c^* {\cal E}^*_0}{1 - \sqrt{2} c \Psi_0 - cc^* {\cal E}^*_0} \: . \ee Notice that, Eqs. (52), (54), (56), (58), (64) and (68) do not coincide to the Kinnersley transformations as they are listed in Esposito and Witten~\cite{18},~\cite{19}, since, in those papers, a different definition for the e/m potentials was used. Agreement is reached, after replacing our $\Psi$ potential by $2 \Psi^*$ (this, of course, is only a redefinition and does not alter the overall concept of the solution generating technique under consideration).

It is worth noting that, the transformation given by Eqs. (64) and (68) was first discovered by Harrison~\cite{14}, who (also) attributed a clear physical interpretation to the complex constant $c$, recognizing that the quantity $H = \frac{1}{2} c c^*$ represents the strength of a uniform magnetic field. Consequently, whenever this transformation is used, it mixes gravity with electromagnetism.

The solution generating techniques considered, have been derived according to a formalism that makes maximal use of a timelike Killing vector. Nevertheless, the Geroch formalism can be successfully carried out, also, upon consideration of spacelike Killing vectors, in a way similar to the above, only provided that some changes on the signs are made.

As an example, we conclude this Section, by deriving an exact solution to the E-M equations that originates from the $\gm_{A}$-solution (see, e.g.,~\cite{20}), upon consideration of a spacelike Killing vector.

In the system of units where both Newton's constant and the velocity of light equal to unity, the $\gm_{A}$-solution in spherical $(t, r, \theta, \phi)$ coordinates reads \bea ds^2 & = & -(2r)^{\gm} \cos^{2 \gm} \left ( \frac{\theta}{2} \right ) d t^2 + \frac{16m^2}{(2r)^{\gm}} \cos^{2 \gm^2 - 2 \gm} \left ( \frac{\theta}{2} \right ) d r^2 \nn \\ & + & \frac{4m^2}{(2r)^{\gm - 2}}  \cos^{2 \gm^2 - 2 \gm} \left ( \frac{\theta}{2} \right ) d \theta^2 + \frac{4m^2}{(2r)^{\gm}} \frac{r^2 \sin^2 \theta}{\cos^{2 \gm} \left ( \frac{\theta}{2} \right )} d \phi^2 \: , \eea where $m$ is the total mass of the central object that is responsible for the gravitational field, and $\gm$ is a constant parameter. Papadopoulos et al.~\cite{20} derived the $\gm_{A}$-solution from its parent one, the $\gm$-solution, using another generation method, named \texttt{limiting procedure for spacetimes} (see, e.g.,~\cite{2}). Quite earlier, Godfrey~\cite{21}, using the fact that all Weyl metrics admit homothetic motion, also reached at the solution given by Eq. (69). A few years later, Lynden-Bell and Pineaut~\cite{22} gave a similar solution, describing a disk of finite radius, while, in~\cite{23}, the same authors discussed realistic rotating disks with frame dragging. On the other hand, Bicak et al.~\cite{24} showed that most of the vacuum Weyl solutions, like the Curzon metric, the $\gm$-metric (both known as Zipoy-Vorhees metrics) and the Israel-Kahn metrics, can be generated by solution (69). In particular, using line sources of finite length, they generated disks corresponding to the Zipoy-Vorhees solution, and discussed several physical properties of such a disc. Furthermore, they showed that, the infinite Lynden-Bell and Pinault solution~\cite{22} can be obtained by taking the upper end of the uniform line-density source to touch the disc, while sending the lower end off to infinity. Finally, Lemos~\cite{25} also used the metric (69) with $\gm = 1$, to discuss the limiting case of an infinite disc, on which the orbital velocity is the velocity of light. In this case, the curved spacetime on each side of the disc is flat, but not on the disc as a whole. In the next Section, we shall explicitly demonstrate the derivation of solutions to the E-M equations, that describe the gravitational field produced by a semi-infinite line source, endowed with either an electric or a magnetic field.

In this Section, we shall use the $\gm_A$-metric in spherical coordinates, given by Eq. (69), in order to generate another exact solution to the E-M equations (cf. Eq. (10) of~\cite{11}), upon consideration of the Kinnersley V transformation based on a spacelike Killing vector. Since the components of the metric tensor (69) are independent of $\phi$, the $\gm_A$-solution admits a well-defined spacelike Killing vector, namely, $\xi_{\phi} = \partial_{\phi}$. In this case, the Ernst potentials associated to the $\gm_A$-metric are \be {\cal E}_0 = -\frac{16 r^2 m^2 \sin^2 \theta}{(2r)^{\gm} \cos^{2 \gm} \left ( \frac{\theta}{2} \right )} ~~~ \mbox{and} ~~~~ \Psi_0 = 0 \: . \ee Performing a Kinnersley V transformation (see, e.g.,~\cite{18},~\cite{19}), we obtain \be {\cal E} = - \frac{16 m^2 r^2 \sin^2 \theta}{(2r)^{\gm} \cos^{2\gm} \left ( \frac{\theta}{2} \right ) + 8 cc^* m^2 r^2 \sin^2 \theta} ~~~ \mbox{and} ~~~~ \Psi = c {\cal E} \: . \ee Accordingly, the resulting solution to the E-M equations is written in the form \bea &&ds^2 = \left [ 1 + \frac{8 cc^* m^2 r^2 \sin^2 \theta}{(2r)^{\gm} \cos^{2 \gm} \left ( \frac{\theta}{2} \right )} \right ]^2 \left [ - (2r)^{\gm} \cos^{2 \gm} \left ( \frac{\theta}{2} \right ) dt^2 + \frac{16m^2}{(2r)^{\gm}} \cos^{2 \gm^2 - 2 \gm} \left ( \frac{\theta}{2} \right ) dr^2 \right . \nn \\ & + & \left . \frac{16m^2}{(2r)^{\gm}} \cos^{2 \gamma^2 - 2 \gamma} \left  ( \frac{\theta}{2} \right ) r^2 d \theta^2 \right ] + \frac{16 m^2}{(2r)^{\gm}} \frac{r^2 \sin^2\theta}{\cos^{2 \gm} \left ( \frac{\theta}{2} \right )} \left [ 1 + \frac{8cc^* m^2 r^2 \sin^2 \theta}{(2r)^{\gm} \cos^{2 \gm} \left ( \frac{\theta}{2} \right )} \right ]^{-2} d \phi^2 . \eea The metric (72) satisfies the Rainich-Misner-Wheeler (RMW) conditions (see, e.g.,~\cite{26}, p. 518), \be {\cal R} = 0 \: , ~~~ {\cal R}_{\nu}^{\mu} {\cal R}_{\sg}^{\nu} = \frac{1}{4} {\cal R}_{\bt}^{\al} {\cal R}_{\al}^{\bt} \dl_{\sg}^{\mu} \ee and \be \al_{\sg ,\ta} - \al_{\ta ,\sg} = 0 \: , \ee where we have set \be \al_{\sg} = \frac{1}{\rh^2} (-g)^{1/2} \varep_{\sg \nu \lm \mu} {\cal R}^{\lm \gm ; \mu} {\cal R}_{\gm}^{\nu} \ee and \be \rh^2 = \frac{1}{4} {\cal R}_{\bt}^{\al} {\cal R}_{\al}^{\bt} \: , \ee and therefore, is an exact solution to the E-M equations. Hence, originating from the $\gm_A$-solution (69) and its spacelike Killing vector $\partial_{\phi}$, upon a Kinnersley V transformation, we have obtained another exact solution to the E-M equations, namely, the magnetized $\gm_A$-metric, given by Eq. (72), which depends on three parameters, i.e., the total mass, $m$, the parameter $\gm$ and the magnetic field strength $H = \frac{1}{2} cc^*$. This solution is, in fact, a member of the broad class of magnetized solutions to the E-M equations, given by Eq. (10) of Richterek et al.~\cite{11}.

\section{Solutions generated from the $\gm_A$-metric in cano-nical coordinates}

The $\gm_A$-solution, given by Eq. (69), is a subsidiary metric of the parent $\gm$-solution, describing the gravitational field of a Weyl source with density $\frac{\gm}{2}$ and length $2m$. The $\gm$-metric exhibits directional behavior (see, e.g.,~\cite{2},~\cite{20}). To better understand this behavior, Papadopoulos et al.~\cite{20} applied the method of limiting procedure for the $\gm$-spacetime, considering the coordinate transformation $r \rarrow (\kp r)^n$ and $t \rarrow \kp^{n (1 - \gm)} t$ (where $\kp$ is a free parameter and $n = 0, 1, 2, ...$), as $\kp \rarrow 0$. Provided that only terms of order $\kp^{n (2 - \gm)}$ are considered, such a transformation actually represents a mapping of the spacetime around a Weyl source with density $\frac{\gm}{2}$ and length $2m$, onto the spacetime around a semi-infinite line source (with the same density) located at the lower half of the $z$-axis. In this context, after a suitable rescaling of the affine parameter, $s$, to get rid of the arbitrary $\kp$ (in this way, however, the mass parameter is also absorbed into $s$, see, e.g.,~\cite{20}), we apply the coordinate transformation $(t, r, \theta, \ph) \rarrow (t, \rh, z, \ph)$, where $(t, \rh, z, \ph)$ are the cylindrical coordinates with $z = r \cos \theta$ and $\rh = r \sin \theta$. Accordingly, the metric (69) results in~\cite{27} \bea ds^2 & = & - \left ( \sqrt{\rh^2 + z^2} + z \right )^{\gm} dt^2 + \left ( \sqrt{\rh^2 + z^2} + z \right )^{\gm (\gm - 1)} \left ( \rh^2 + z^2 \right )^{- \gm^2} \left [ d \rh^2 + dz^2 \right ] \nn \\ & + & \rh^2 \left ( \sqrt{\rh^2 + z^2} + z \right )^{- \gm} d \phi^2 \: .\eea Verdaguer~\cite{28} showed that the spacetime (77) corresponds to an $n$-soliton solution to the Einstein field equations. Now, setting \be \nu = \frac{\gm}{2} \ln \left [ \sqrt{\rh^2 + z^2} + z \right ] ~~\mbox{and}~~ \mu = \frac{\gm^2}{2} \ln \frac{\left [ \sqrt{\rh^2 + z^2} + z \right ]}{\sqrt{\rh^2 + z^2}} \: , \ee which satisfy the relations \be \nu_{, \rh \rh} + \frac{\nu_{, \rh}}{\rh} + \nu_{, z z} = 0 \: , \ee and \be \mu_{, \rh} = \rh \left ( \nu_{,\rh}^2 - \nu_{,z}^2 \right ) \: ,~~~ \mu_{, z }= 2 \rh \nu_{, \rh} \nu_{, z} \: , \ee the line element (77) is written in a Weyl canonical form, i.e., \be ds^2 = - e^{2 \nu} dt^2 + e^{- 2 \nu} \left [ e^{2 \mu} \left ( d \rh^2 + dz^2 \right ) + \rh^2 d \ph^2 \right ] \: . \ee From this point of view, and as long as $\gm \neq 0, 1$, the solution (77) does describe the gravitational field around a semi-infinite line source, with uniform density equal to $\frac{\gm}{2}$, located at the lower half of the $z$-axis, i.e., on $z \leq 0$ (see, e.g.,~\cite{2},~\cite{29}), thus explaining why the metric (77) is not reflection symmetric.\footnote{For $\gm \neq 0, 1$, the metric (77) is a $\gm$-soliton, generated by a Euclidean metric~\cite{27},~\cite{28}.}

\subsection{Solutions generated by the action of a timelike Killing vector}

A timelike Killing vector of the (seed) $\gm_A$-metric (77) is \be \xi^{\mu} = [1, 0, 0, 0] \: , \ee for which, we have \be \xi^{\mu} \xi_{\mu} = - \lm_0^t = - \left ( \sqrt{\rh^2 + z^2} + z \right )^{\gm} = g_{tt} \: , \ee while both the twist vector, $\om_0^{\al}$, and the associated potential, $\psi_0$, are equal to zero. In this case, the original Ernst potentials are given by \be {\cal E}^t_0 = \lm_0^t = \left ( \sqrt{\rh^2 + z^2} + z \right )^{\gm} ~~\mbox{and}~~~ \Psi^t_0 = 0 \: , \ee where \texttt{"t"} stands for \texttt{"timelike"}. Upon consideration of a Kinnersley V (or Harrison's) transformation, the new Ernst potentials, ${\cal E}^t$ and $\Psi^t$, are obtained by the formulae \be {\cal E}^t = \frac{{\cal E}^t_0}{1 - c (\Psi^t_0)^* - \frac{1}{2} cc^* {\cal E}^t_0} ~~\mbox{and}~~ \Psi^t = \frac{\Psi^t_0 + c {\cal E}^t_0}{1 - c^* \Psi^t_0 - \frac{1}{2} cc^* {\cal E}^t_0} \ee (see, e.g.,~\cite{18},~\cite{19}). Now, in view of Eqs. (84), Eqs. (85) yield \be {\cal E}^t = \frac{\left ( \sqrt{\rh^2 + z^2} + z \right )^{\gm}}{1 - \frac{1}{2} cc^* \left ( \sqrt{\rh^2 + z^2} + z \right )^{\gm}} \: , \ee and \be \Psi^t = \frac{c \left ( \sqrt{\rh^2 + z^2} + z \right )^{\gm}}{1 - \frac{1}{2} cc^* \left ( \sqrt{\rh^2 + z^2} + z \right )^{\gm}} = c {\cal E}^t \: . \ee Furthermore, by virtue of Eq. (35), we have \be {\cal E}^t + \frac{1}{2} \Psi^t (\Psi^t)^* = \lm^t + 2 \imath \psi^t \: , \ee which, in view of Eq. (87) and the fact that ${\cal E}^t$ is real, leads to \be \lm^t = {\cal E}^t + \frac{1}{2} cc^* \left ( {\cal E}^t \right )^2 = \frac{ \left ( \sqrt{\rh^2 + z^2} + z \right )^{\gm}}{ \left [ 1 - \frac{1}{2} cc^* \left ( \sqrt{\rh^2 + z^2} + z \right )^{\gm} \right ]^2} ~~ \mbox{and}~~~ \psi^t = 0 \: . \ee The corresponding twist vector, $\om_a$, is obtained by the formula \be \tilde{D}_a \psi^t = \om^t_a + \frac{\imath}{4} \left [ \Psi^t \tilde{D}_a \left ( \Psi^t \right )^* - \left ( \Psi^t \right )^* \tilde{D}_a \Psi^t \right ] \: . \ee Once again, in view of Eq. (87) and the fact that ${\cal E}^t$ is real, Eq. (90) results in \be \Psi^t \tilde{D}_a (\Psi^t)^* - (\Psi^t)^* \tilde{D}_a (\Psi^t) = 0 \: . \ee Finally, with the aid of Eqs. (89) and (90), we conclude that \be \om^t_a = 0 \: . \ee

Based on the technique of generating solutions to the E-M equations presented in Section 3, upon consideration of the new Ernst potentials, ${\cal E}^t$ and $\Psi^t$, the twist vector $\om^t_a$ and the associated potential $\psi^t$, we may now determine the line element of an exact solution to the E-M equations, that is generated by the $\gm_A$-metric (77) upon the action of a timelike Killing vector of norm $\lm^t$. To do so, first, we express Eq. (77) in the (more convenient) form \bea ds^2 & = & - \left ( \sqrt{\rh^2 + z^2} + z \right )^{\gm} dt^2 \\ & + & \frac{1}{\left ( \sqrt{\rh^2 + z^2} + z \right )^{\gm}} \cdot \left [ \frac{\left ( \sqrt{\rh^2 + z^2} +z \right )^{\gm^2}}{\left ( \sqrt{\rh^2 + z^2} \right )^{\gm^2}} d \rh^2 + \rh^2 d \phi^2 + \frac{\left ( \sqrt{\rh^2 + z^2} + z \right )^{\gm^2}}{\left ( \sqrt{\rh^2 + z^2} \right )^{\gm^2}} dz^2 \right ] \: . \nn \eea Accordingly, the solution generated by solution (77) (or, equivalently, by solution (93)) reads \bea && ds^2 = - \frac{\left ( \sqrt {\rh^2 + z^2} + z \right )^{\gm}}{\left [ 1 - \frac{1}{2} cc^* \left ( \sqrt{\rh^2 + z^2} + z \right )^{\gm} \right ]^2} dt^2 \\ && + \frac{ \left [ 1 - \frac{1}{2} cc^* \left ( \sqrt{\rh^2 + z^2} + z \right )^{\gm} \right ]^2}{ \left ( \sqrt {\rh^2 + z^2} + z \right )^{\gm}} \cdot \left \lbrace \frac{ \left ( \sqrt{\rh^2 + z^2} + z \right )^{\gm^2}}{ \left ( \sqrt{\rh^2 + z^2} \right )^{\gm^2}} \left [ d \rh^2 + d z^2 \right ] + \rh^2 d \phi^2 \right \rbrace \: . \nn \eea The metric given by Eq. (94) also satisfies the Rainich-Misner-Wheeler conditions (73) - (76), and therefore, is an exact solution to the E-M equations. Recall that, according to Petrov's classification~\cite{7}, since ${\cal E}^t$ is real and $\Psi^t$ is complex, solution (94) describes a stationary e/m field. In fact, this solution is identical to Eq. (31) of Richterek et al.~\cite{11}, which describes the curved spacetime around a semi-infinite line source endowed with an electric field.

\subsection{Solutions generated by the action of a spacelike Killing vector}

Now, let us begin with a spacelike Killing vector of the (seed) $\gm_A$-metric, namely, \be \xi^{\nu} = [0, 0, 0, 1] \ee where \be \xi^{\nu} \xi_{\nu} = - \lm_0^s = \frac{\rh^2}{\left ( \sqrt{\rh^2 + z^2} + z \right )^{\gm}} = g_{\phi \phi} \: . \ee As we have already noted, as regards the $\gm_A$-metric, both the twist vector $\om_0^{\al}$ and the associated potential $\psi_0$ are equal to zero. In the case of a spacelike Killing vector, the corresponding Ernst potentials are given by \be {\cal E}^s_0 = \lm_0^s = - \frac{\rh^2}{ \left ( \sqrt{\rh^2 + z^2} + z \right )^{\gm}} ~~\mbox{and}~~~  \Psi^s_0 = 0 \: , \ee where \texttt{"s"} stands for \texttt{"spacelike"}. In this case, upon consideration of a Kinnersley V transformation, the new Ernst potentials, ${\cal E}^s$ and $\Psi^s$, are given by the formulae \be {\cal E}^s = \frac{{\cal E}^s_0}{1 - c ( \Psi^s_0)^* - \frac{1}{2} cc^* {\cal E}^s_0} ~~\mbox{and}~~ \Psi^s = \frac{\Psi^s_0 + c {\cal E}^s_0}{1 - c^* \Psi^s_0 - \frac{1}{2} cc^* {\cal E}^s_0} \ee (see, e.g.,~\cite{18},~\cite{19}). By virtue of Eqs. (97) and (98), we obtain \be {\cal E}^s = \frac{- \rh^2}{ \left ( \sqrt{\rh^2 + z^2} + z \right )^{\gm} + \frac{1}{2} cc^* \rh^2} \ee and \be \Psi^s = \frac{- c \rh^2}{ \left ( \sqrt{\rh^2 + z^2} + z \right )^{\gm} + \frac{1}{2} cc^* \rh^2} = c {\cal E}^s \: . \ee In a fashion analogous to the procedure described in Section 4.1, we end up with the following results \be \lm^s = {\cal E}^s + \frac{1}{2} cc^* \left ( {\cal E}^s \right )^2 = - \frac{\rh^2 \left ( \sqrt{\rh^2 + z^2} + z \right )^{\gm}}{ \left [ \left ( \sqrt{\rh^2 + z^2} + z \right )^{\gm} + \frac{1}{2} cc^* \rh^2 \right ]^2} \ee and \be \psi^s = 0 = \om^s_a \: . \ee Once again, based on the technique of generating solutions to the E-M equations presented in Section 3, upon consideration of the new Ernst potentials ${\cal E}^s$ and $\Psi^s$, the twist vector $\om^s_a$, and the associated potential $\psi^s$, we can determine the line element of another exact solution to the E-M equations, that is generated by the $\gm_A$-metric upon the action of a spacelike Killing vector of norm $\lm^s$. To do so, first, we rewrite Eq. (77) in the form \bea ds^2 & = & - \frac{\rh^2}{\left ( \sqrt{\rh^2 + z^2} + z \right )^{\gm}} \left ( - d \phi^2 \right ) \\ & - & \frac{\left ( \sqrt{\rh^2 + z^2} + z \right )^{\gm}}{\rh^2} \left [ \rh^2 dt^2 - \frac{\rh^2 \left ( \sqrt{\rh^2 + z^2} + z \right )^{\gm (\gm - 2)}}{\left ( \sqrt{\rh^2 + z^2} \right )^{\gm^2}} d \rh^2 - \frac{\rh^2 \left ( \sqrt{\rh^2 + z^2} + z \right )^{\gm (\gm - 2)}}{\left ( \sqrt{\rh^2 + z^2} \right )^{\gm^2}} dz^2 \right ] \: , \nn \eea in view of which, the resulting solution reads \bea ds^2 & = & - \frac{ \left [ \left ( \sqrt {\rh^2 + z^2} + z \right )^{\gm} + \frac{1}{2} cc^* \rh^2 \right ]^2}{ \left ( \sqrt{\rh^2 + z^2} + z \right )^{\gm}} dt^2 + \nn \\ & + & \frac{ \left [ \left ( \sqrt {\rh^2 + z^2} + z \right )^{\gm} + \frac{1}{2} cc^* \rh^2 \right ]^2 \left ( \sqrt{\rho^2 + z^2} + z \right )^{\gm (\gm-3)}}{ \left ( \sqrt{\rh^2 + z^2} \right )^{\gm^2}} \left [ d \rh^2 + dz^2 \right ] + \nn \\ & + & \frac{\rh^2 \left ( \sqrt {\rh^2 + z^2} + z \right )^{\gm}}{ \left [ \left ( \sqrt {\rh^2 + z^2} + z \right )^{\gm} + \frac{1}{2} cc^* \rh^2 \right ]^2} d \phi^2 \: . \eea Once again, the metric (104) satisfies the Rainich-Misner-Wheeler conditions (73) - (76) and, therefore, is an exact solution to the E-M equations. In addition, since ${\cal E}^s$ is real and $\Psi^s$ is complex, according to Petrov's classification~\cite{7}, solution (104) also describes a stationary e/m field. In fact, this solution is identical to Eq. (32) of Richterek et al.~\cite{11}, which describes the curved spacetime around a semi-infinite line source endowed with a magnetic field.

Finally, we should stress that, the continuous transformations given by Eqs. (52), (54), (56), (58), (62) and (68) are not the only ones that leave the Lagrangian (45) invariant under the variation of the generalized coordinates of the potential space. There are also other transformations, which lead to solutions of the Einstein or/and the E-M equations other than those derived in the present article. Some of those transformations are presented in the Appendix C.

\section{Discussion and Conclusions}

In the present article, we assume that the curved spacetime $(M, g_{\al \bt})$ admits symmetries, generated by Killing vectors. This assumption allow us to factorize the four-dimensional manifold, yielding a three-dimensional space, $S$, that is constructed by the trajectories of the Killing vector $\xi^{\mu}$. Accordingly, we attribute a Riemannian structure on $S$, with metric $h_{ab} = g_{ab} + \lm^{-1} \xi_a \xi_b$, where $\lm$ is the norm of the Killing vector under consideration.

The technique of generating solutions to the Einstein field equations, rests in decomposing them in terms of quantities of the quotient manifold, $S$, that are associated to $\xi^{\mu}$, such as the gradient of its norm, $D_a \lm$, and the twist vector, $\om_a \equiv D_a \om$, i.e., the gradient of the twist potential. In this case, the principal variables attributed to a stationary solution of the Einstein equations are $h_{ab}$, $\lm$ and $\om$. Geroch~\cite{1} determined transformations that relate $\left \lbrace h_{ab}, \lm, \om \right \rbrace$ to a new set, $\left \lbrace h_{ab}, \lm^{\prime}, \om^{\prime} \right \rbrace$, which continues to solve the Einstein equations. In other words, given the principle variables of a stationary solution to the Einstein field equations $\left \lbrace h_{ab}, \lm, \om \right \rbrace$, we may construct the corresponding quantities of another solution $\left \lbrace h_{ab}, \lm^{\prime}, \om^{\prime} \right \rbrace$, and, through them, determine the associated metric, $g_{\al \bt}^{\prime}$. In this article, we applied this method to the stationary E-M equations (see also \cite{6},~\cite{13} and~\cite{16}).

In this case, along with the various quantities associated to the Killing vector, we also have the decomposition of the Faraday tensor into an electric, $E_{\al}$, and a magnetic part, $B_{\al}$. The Maxwell equations on the quotient manifold suggest that both quantities are due to a potential, i.e., $E_a = D_a \eps$ and $B_a = D_a \bt$. All the stationary E-M equation data (i.e., $h_{ab}$, $\lm$, $\om$, $\eps$ and $\bt$) are accordingly expressed in terms of the Ernst potentials, ${\cal E}$ and $\Psi$ (cf. Eqs. (32) and (35)). Now, the basic quantities concerning the original set of the E-M equations are $\left \lbrace h_{ab}, {\cal E}_0, \Psi_0 \right \rbrace$, and the solution generating technique under study involves the determination of transformations, which, originating from $\left \lbrace h_{ab}, {\cal E}_0, \Psi_0 \right \rbrace$, produce a new set $\left \lbrace h_{ab}, {\cal E}, \Psi \right \rbrace$, that still satisfies the E-M equations. Given the complexity of these equations, it is rather surprising that such transformations actually exist.

Notice that, in view of the aforementioned technique, if someone begins with a static solution, the resulting solution is stationary. Furthermore, application of this technique adds twist (i.e., rotation) to an already existing solution. Finally, it is not necessary to begin with a timelike Killing vector; one could also make use of the solutions generating technique under study, upon consideration of a spacelike Killing vector.

Our quest for exact solutions to the E-M equations rests upon the $\gm_A$-solution as the seed metric. In spherical coordinates, this metric admits a well defined spacelike Killing vector, $\partial_{\phi}$. Accordingly, using the Kinnersley V transformation, which, in fact, is Harrison's transformation that has the ability of mixing gravity with electromagnetism~\cite{14}, we derive another three-parameter solution to the E-M equations, given by Eq. (72). These three parameters are the total mass of the central object responsible for the gravitational field, $m$, the parameter $\gm$, and the strength of a uniform magnetic field, $H = \frac{1}{2} cc^*$. Solution (72) belongs to the broad class of solutions to the E-M equations given by Eq. (10) of~\cite{11}. For $c = 0$, the metric (72) reduces to a solution, the astrophysical implications of which have been extensively discussed in~\cite{24}, and its mathematical properties in~\cite{27} and~\cite{28}.

Furthermore, expressing the $\gm_A$-solution in canonical coordinates and performing a Kinnersley V transformation, we have explicitly demonstrated the way of generating exact solutions to the E-M equations, either by the use of a timelike Killing vector or by the use of a spacelike Killing vector, namely, Eqs. (94) and (104), respectively. At this point, we cannot help but noticing that, as far as the $\gm_A$-metric is concerned, implementation of a Kinnersley V transformation upon the action of a timelike Killing vector leads to solution (94), which is endowed with an electric field, while the corresponding transformation upon the action of a spacelike Killing vector leads to solution (104), which is endowed with a magnetic field (cf. Eqs. (31) and (32) of~\cite{11}, respectively). To which extend is this a property of the $\gm_A$-solution alone or a general rule of the E-M theory itself - i.e., a Kinnersley V transformation with timelike (spacelike) Killing vector always leads to solutions with electric (magnetic) field - is an intriguing question, which we will attempt to address in a future work.

An alternative way of finding new solutions to the E-M equations, is to begin with the action for the stationary E-M equations, in other words, the Einstein-Hilbert action supplemented by the action for the e/m field. The existence of a Killing vector factorizes this action in the form \be I \sim \int d^4 x \sqrt{h} \left [ {\cal R} (h_{ab}) + g_{AB} \Phi_{, a}^A {\Phi^B}^{, a} \right ] \ee (see, e.g.,~\cite{6},~\cite{30}). Clearly, the first term of Eq. (105) is the standard geometric action for the gravitational field on the three-dimensional space constructed by the trajectories of the Killing vectors $(S, h_{ab})$, while, the second term may be considered as representing a non-linear $\sg$-model (analogous to the corresponding quantum-mechanical models) with metric, $g_{AB}$, determined exclusively by the E-M equations. In this case, the potentials ${\cal E}$, $\Psi$, ${\cal E}^{*}$, and $\Psi^{*}$ correspond to the (generalized) coordinates of this space. We refer to this manifold as the \texttt{"moduli space"} $(N, g_{AB})$ of the stationary E-M equations. Any solution to the stationary E-M equations represents an extremum of the action (105) on the moduli space. It is now straightforward to describe the real nature of the solution generation technique.

What we really do is, to find transformations in the moduli space that keep us at the extremum of the action (105), and this can be achieved only by determining the Killing vectors of the moduli space. In other words, integration of the Killing equations on the moduli space, provides a set of smooth transformations which conserve the extremum, i.e., a set of transformations among the possible solutions to the stationary E-M equations. The moduli space $(N, g_{AB})$ is, in fact, a Kaehler manifold, on which, for each solution, the set of allowed transformations is the isotropy group of the extremum~\cite{30}. For this reason, the transformations originating from the Killing trajectories on the moduli space do not exhaust the set of transformations that preserve a Kaehler extremum; clearly, the isotropy group is larger than that originating from the Killing vectors alone. This is not yet quite understood, and therefore, further analysis on the geometry of the moduli space is needed.

At this point, we should stress that, the solution generating technique described in this article, is not necessarily restricted to four dimensions, neither is restricted to abelian fields, nor is restricted to timelike Killing vectors of the original four-dimensional spacetime. For instance, if we begin with a stationary, axi-symmetric solution, we may use the (solution-generating) transformation based on a timelike Killing vector, as well as on a spacelike Killing vector. It is known that, these two transformations do not commute, but still, the situation is not well understood. On the other hand, the case of non-abelian fields is simply not known.

Finally, it is very interesting that, the part of the action (105) which involves the potentials has the form of a non-linear $\sg$-model. From this point of view, it appears that, there is a harmonic mapping from $(S, h_{ab})$ to $(N, g_{AB})$, i.e., the E-M equations for the potentials may be viewed as geodesic equations on the moduli space and vice versa. The theoretical explanation of such an observation would be of great importance, since it could lead to a unified description of the majority of the solution generating techniques, and it will be the scope of a future work.

\vspace{.5cm}

\textbf{Acknowledgments:} The authors would like to thank the anonymous referee for his/her kind comments and his/her useful suggestions, which greatly improved this article's final form. This work has been supported by the General Secretariat for Research \& Technology of Greece and by the European Social Fund, within the framework of the action "EXCELLENCE".

\section*{Appendix A: The wave operator on the four-dimensional manifold $M$}

To determine the d' Alembertian of the norm of a Killing vector on $M$, we begin with $$ \nabla_{\al} \nabla^{\al} \xi^{\gm} = - {\cal R}^{\gm}_{\mu} \xi^{\mu} \: , \eqno{(A1)} $$ hence, $$ \xi_{\gm} \nabla^{\al} \nabla_{\al} \xi^{\gm} = \nabla_{\al} \left ( \xi^{\gm} \nabla^{\al} \xi_{\gm} \right ) - \left ( \nabla_{\al} \xi_{\gm} \right ) \left ( \nabla^{\al} \xi^{\gm} \right ) = - {\cal R}^{\gm \mu} \xi_{\gm} \xi_{\mu} \eqno{(A2)} $$ or, else, $$ -\frac{1}{2} \nabla_{\al} \nabla^{\al} \lm - \left ( \nabla_{\al} \xi_{\gm} \right ) \left ( \nabla^{\al} \xi^{\gm} \right ) = - {\cal R}^{\gm \mu} \xi_{\gm} \xi_{\mu} \: . \eqno{(A3)} $$ However, $$ \left ( \nabla_{\al} \xi_{\gm} \right ) \left ( \nabla^{\al} \xi^{\gm} \right ) = \lm ^{-1} \left [ 2 \om^{\al} \om_{\al} - \frac{1}{2} \left ( \nabla_{\al} \lm \right ) \left ( \nabla^{\al} \lm \right ) \right ] \eqno{(A4)} $$ and therefore, $$ \nabla^{\al} \nabla_{\al} \lm = \lm^{-1} \left ( \nabla_{\al} \lm \right ) \left ( \nabla^{\al} \lm \right ) - 4 \lm^{-1} \om^{\al} \om_{\al} + 2 {\cal R}_{\al \bt} \xi^{\al} \xi^{\bt} \: , \eqno{(A5)} $$ which coincides to Eq. (15).

\section*{Appendix B: The Ricci tensor of the three-dimensional manifold $S$}

Let $k^a$ be a vector of $S$, such that $k^a \xi_a = 0 = {\cal L}_{\xi} k^a$. Then, by definition of the curvature tensor on $S$, we have $$ \frac{1}{2} {\cal R}^{(S)}_{abcd} k^d = D_{[a}D_{b]} k _c = h^m_{[a}h^n_{b]} h^p_c \nabla_m \left ( h^k_n h^s_p \nabla_r k_s \right ) \: . \eqno{(B1)} $$ In view of Eq. (B1), we deduce that, the Riemann tensor on $S$, in terms of the corresponding quantity of $M$, is given by $$ {\cal R}^{(S)}_{abcd} = h^m_{[a}h^n_{b]} h^k_{[c}h^l_{d]} \: \left [ {\cal R}^{(M)}_{m n k l} - 2 \lm^{-1} \left ( \nabla_m \xi_n \right ) \left ( \nabla_k \xi_l \right ) - 2 \lm^{-1} \left ( \nabla_m \xi_k \right ) \left ( \nabla_n \xi_l \right ) \right ] \: . \eqno{(B2)} $$ Accordingly, the Ricci tensor of the three-dimensional quotient manifold, $S$, reads $$ {\cal R}^{(S)}_{bd} = h^{ac} {\cal R}^{(S)}_{abcd} = h^{pr} h^q_b h^s_d {\cal R}^{(M)}_{pqrs} - 2 \lm^{-1} h^p_{[a}h^q_{b]} h^{qr} h^s_d \left [ \left ( \nabla_p \xi_q \right ) \left ( \nabla_r \xi_s \right ) + \left ( \nabla_p \xi_r \right ) \left ( \nabla_q \xi_s \right ) \right ] \: . \eqno{(B3)} $$

\section*{Appendix C: Other continuous transformations}

Upon consideration of a static vacuum solution $(M, g_{\al \bt})$ to the Einstein field equations, that admits a timelike Killing vector, $\xi^{\mu}$, the Lagrangian (45), in terms of quantities on the three-dimensional slice $(S, h_{ab})$ perpendicular to $\xi^{\mu}$, is written in the form $$ L = \sqrt{\tilde{h}} \: \left [ \tilde{\cal R}^{(S)} - \frac{\lm^{-2}}{2} \tilde{h}^{ab} \left ( \tilde {D} \lm \right ) \left ( \tilde{D} \lm \right ) \right ] \: . \eqno{(C1)} $$ Clearly, the Lagrangian (C1) remains invariant under the transformation $$ \lm \rightarrow \lm^{-1} \: . \eqno{(C2)} $$ This, of course, means that, also, the equations $$ \lm \tilde{D}^{a} \tilde{D}_{a} \lm = \tilde{h}^{ab} \left ( \tilde {D}_{a} \lm \right ) \left ( \tilde{D}_{b} \lm \right ) \eqno{(C3)} $$ and $$ 2 \lm^2 \tilde{\cal R}^{(S)}_{ab} = \tilde{D}_{(a} \lm \tilde{D}_{b)} \lm \: , \eqno{(C4)} $$ remain invariant under the transformation (C2). This is the Buchdahl~\cite{31} transformation.

There is another transformation, found by Demianski (see, e.g.,~\cite{2}), which can be described in a relatively simple manner. A stationary vacuum solution satisfies Eqs. (41) and (42) with $\Psi = 0$, i.e., $$ \lm \tilde{D}^a \tilde{D}_a {\cal E} = \left ( \tilde{D}^a {\cal E} \right ) \left ( \tilde{D}_a {\cal E} \right ) \: , \eqno{(C5)} $$ and $$ 2 \lm^2 \tilde{\cal R}^{(S)}_{ab} = \tilde{D}_{(a} {\cal E} \tilde{D}_{b)} {\cal E}^* \: . \eqno{(C6)} $$ Accordingly, defining as $$ \Psi^{\prime} = c \: \frac{{\cal E} + 1}{{\cal E} - 1} \: , \eqno{(C7)} $$ and $$ \lm^{\prime} = \frac{1}{2} \left ( \Psi^{\prime} {\Psi^*}^{\prime} - c^2 \right ) \: , \eqno{(C8)} $$ it is straightforward to show that $\Psi^{\prime}$ and $\lm^{\prime}$ satisfy the equations $$ \lm^{\prime} \tilde{D}^a \tilde{D}_a \Psi^{\prime} = {\Psi^{\prime}}^* \left ( D^a \Psi^{\prime} \right ) \left ( \tilde{D}_a \Psi^{\prime} \right ) \: , \eqno{(C9)} $$ and $$ 2 \left ( \lm^{\prime} \right )^2 \tilde{\cal R}^{(S)}_{ab} = - \left ( {\cal E}^{\prime} + {{\cal E}^{\prime}}^* \right ) \tilde{D}_{(a} \Psi^{\prime} \tilde{D}_{b)} {\Psi^{\prime}}^* \: . \eqno{(C10)} $$ In this case, provided that Eqs. (C9) and (C10) are satisfied for ${\cal E}^{\prime} = constant$, Eqs. (41) and (42) are also satisfied. Hence Eqs. (C7) and (C8) demonstrate the way to transform a stationary vacuum solution, ${\cal E}$, to a stationary electrovacuum solution, with ${\cal E}^{\prime} = constant$.

Another transformation, from a stationary axisymmetric vacuum to a static electrovacuum, was found by Bonnor~\cite{32},~\cite{33}. This transformation can be described upon the observation that the Langrangian for a stationary vacuum, $$ L = \sqrt{\tilde{h}} \: \left \lbrace \tilde{\cal R}^{(S)} - \frac{1}{2} \lm^{-2} \tilde{h}^{ab} \left [ \left ( \tilde{D}_a \lm \right ) \left ( \tilde{D}_b \lm \right ) + 4 \left ( \tilde{D}_a \Psi \right ) \left ( \tilde{D}_b \Psi \right ) \right ] \right \rbrace \: , \eqno{(C11)} $$ upon the transformation $$ \lm^2 = \lm^{\prime} ~~~\mbox{and}~~~~ \Psi = \imath 8^{-1/2} \eps \: , \eqno{(C12)} $$ results in $$ L = \sqrt{\tilde{h}} \: \left \lbrace \tilde{\cal R}^{(S)} - \frac{1}{8} {\lm^{\prime}}^{-2} \tilde{h}^{ab} \left [ \left ( \tilde{D}_a \lm^{\prime} \right ) \left ( \tilde{D}_b \lm^{\prime} \right ) - 2 \lm^{\prime} \left ( \tilde{D}_a \eps \right ) \left ( \tilde{D}_b \eps \right ) \right ] \right \rbrace \: . \eqno{(C13)} $$ For axisymmetric solutions, the Euler equations originating from Lagrangian (C13), after a rearrangement of constants, are the field equations that determine static axisymmetric electrovacuum metrics.


\begin{thebibliography}{35}

\bibitem{1} Geroch, R., J. Math. Phys. {\bf 12}, 918 (1971); {\bf 13}, 394 (1972)
\bibitem{2} Stephani, H., Kramer, D., Mac Callum, M., Hoenselaers, C., and Herlt, E., \emph{Exact Solutions to the Einstein's Field Equations: Second Edition}, Cambridge University Press, Cambridge (2003)
\bibitem{3} Ehlers, J., and Kundt, W., \emph{Gravitation (Ch. 2): An Introduction to Current Research}, L. Witten (ed.) Wiley, NY (1962)
\bibitem{4} Kramer, D., Neugebauer, G., and Stephani, H., Forts. der Phys. {\bf 20}, 1 (1972)
\bibitem{5} Kinnersley, W., J. Math. Phys. {\bf 14}, 651 (1973)
\bibitem{6} Kinnersley, W., J. Math. Phys. {\bf 18}, 1529 (1976)
\bibitem{7} Petrov, A. Z., \emph{Einstein Spaces}, Pergamon Press, NY (1969)
\bibitem{8} Manko, V. S., Mielke, E. W., Sanabria-Gomez, J. D., Phys. Rev. {\bf D 61}, 081501 (2000); Manko, V. S., Marmo, G., Porzio, A., Solimeno, S., Zaccaria, F., Phys. Rev. {\bf D 62}, 044048 (2000)
\bibitem{9} Bicak, J., \emph{Einstein Field Equations and their Physical Implications (Selected essays in honnor of J.Ehlers)}, Lecture Notes in Physics {\bf 540}, B. G. Schmidt (ed.) Springer-Verlag, Berlin (2000), \texttt{arXiv: 0004016 [gr-qc]}
\bibitem{10} Richterek, L., Novotn{\'y}, J., and Horsk{\'y}, J., CzJPh {\bf 50}, 925 (2000), \texttt{arXiv: 0003004 [gr-qc]}
\bibitem{11} Richterek, L., Novotn{\'y}, J., and Horsk{\'y}, J., CzJPh {\bf 52}, 1021 (2002), \texttt{arXiv: 0209094 [gr-qc]}
\bibitem{12} Berti, E., and Stergioulas, N., MNRAS {\bf 350}, 1416 (2004)
\bibitem{13} Neugebauer, G., and Kramer, D., Ann. der Phys. {\bf 29}, 62 (1969)
\bibitem{14} Harrison, B. K., J. Math. Phys. {\bf 9}, 1744 (1968)
\bibitem{15} Contopoulos, I. G., Esposito, F. P., Kleidis, K., Papadopoulos D. B., and Witten, L., \textit{Generating solutions to the Einstein field equations}, IJMP {\bf D} \textit{submitted} (2015) \texttt{arXiv: 1501.03968v3 [gr-qc]}
\bibitem{16} Ernst, F. J., Phys. Rev. {\bf 167}, 1175 (1968); {\bf 168}, 1415 (1968); J. Math. Phys. {\bf 14}, 651 (1973); J. Math. Phys. {\bf 17}, 45 (1976)
\bibitem{17} Eisenhart, L., \emph{Riemannian Geometry}, Princeton (1949)
\bibitem{18} Esposito F. P., and Witten, L., Phys. Rev. {\bf D8}, 3302 (1973)
\bibitem{19} Esposito F. P., and Witten, L., Gen. Relativ. Gravit. {\bf 6}, 387 (1975); Witten, L. \emph{The generation of solutions to the Einstein-Maxwell equations}, in Proceedings of Second Latin American Conference of Relativity and Gravitation, Caracas, Venezuela (1976)
\bibitem{20} Papadopoulos, D., Stewart B., and Witten, L., Phys. Rev. {\bf D24}, 320 (1981); and references therein
\bibitem{21} Godfrey, B. B., Gen. Relativ. Gravit. {\bf 3}, 3 (1972)
\bibitem{22} Lynden-Bell, D., and Pineault, S., MNRAS {\bf 185}, 679 (1978)
\bibitem{23} Lynden-Bell, D., and Pineault, S., MNRAS {\bf 185}, 695 (1978)
\bibitem{24} Bicak, J., Lynden-Bell, D., and Katz,J., Phys. Rev. {\bf D47}, 4334 (1993)
\bibitem{25} Lemos, J. P. S., MNRAS {\bf 230}, 451 (1998)
\bibitem{26} Adler, R., Bazin, M., and Sciffer, M., \emph{Introduction to General Relativity}, McGraw-Hill, New York (1975)
\bibitem{27} Papadopoulos, D., Nuovo Cimento {\bf 44}, 497 (1985)
\bibitem{28} Verdaguer, E., J. Phys. {\bf A 15}, 1261 (1981)
\bibitem{29} Synge, J. L., \emph{Relativity and General Theory}, North Holland, Amsterdam (1966), p. 312
\bibitem{30} Argyres, P. C., Awad, A., Braun, G., and Esposito, F.P., JHEP {\bf 07}, 060 (2003)
\bibitem{31} Buchdahl, H. A., Quant. J. Math. Ox. {\bf 5}, 1161 (1954)
\bibitem{32} Bonnor, W. B., Z. Phys. {\bf 161}, 439 (1961)
\bibitem{33} Bonnor, W. B., Z. Phys. {\bf 190}, 444 (1966)

\end{thebibliography}
\end{document}